\documentclass[preprint,showkeys,secnumarabic,amsfonts,showpacs,amsmath,amssymb]{revtex4}

\usepackage[dvips]{color}
\usepackage{epsfig}
\usepackage{amsmath}
\usepackage{graphicx}
\date{\today}

\begin{document}

\title{Interacting Cosmic Fluids in Brans-Dicke Cosmology }

\author{Hossein Farajollahi}
\email{hosseinf@guilan.ac.ir} \affiliation{Department of Physics,
University of Guilan, Rasht, Iran}

\author{Narges Mohamadi}
\email{nmohamadi@guilan.ac.ir}\affiliation{Department of Physics,
University of Guilan, Rasht, Iran}

\author{Hamed Amiri}
\email{hamiri@guilan.ac.ir}
\affiliation{Department of Physics,
University of Guilan, Rasht, Iran}

\begin{abstract}

We provide a detailed description for power-law scaling FRW
cosmological models in Brans-Dicke theory dominated by two
interacting fluid components during the expansion of the universe.

\end{abstract}

\keywords{Brans-Dicke cosmology; scalar field; expansion; inflation; dark energy; phantom; interaction}

\maketitle

\section{Introduction}

Brans-Dicke (BD) theory is considered as a natural extension of
Einstein' s general theory of relativity \cite{Brans}, where the gravitational constant becomes time dependent varying as inverse
of a time dependent scalar field which couples to gravity with a
coupling parameter $\omega$. Many of the cosmological problems
\cite{Steinhardt}--\cite{Singh} can be
successfully explained by using this theory. One important property of BD theory is
that it gives simple expanding solutions \cite{Mathiazhagan}\cite{La} for
the scalar field $\varphi(t)$ and the scale factor $a(t)$ which are
compatible with solar system observations
\cite{Perlmutter}--\cite{Garnavich}, in which impose lower bound on $\omega$ ($|\omega| \geq
10^4$) \cite{Bertotti}.

On the other hand, recent observational data give a strong motivation to study general
properties of Friedmann-Robertson-Walker (FRW) cosmological models
containing more than one fluid ( for example, see \cite{Cataldo}). Usually the universe is modeled with
perfect fluids and with mixtures of non interacting perfect fluids.
However there are no observational data confirming that this is the
only possible scenario. This means that we can consider plausible
cosmological models containing fluids which interact with each
other.

In this work, we follow the authors in \cite{Cataldo} but apply the two fluid interaction in
 FRW BD cosmologies. In this case the transfers of energy
among these fluids in relation to the BD scalar field play an important role in the formalism. There are many cosmological
situations with the exchange of energy that will be investigated in this work. Although
the interaction between for example dust-like matter and radiation was previously
considered in \cite{Cataldo}, \cite{Wood} and \cite{Gunzing}, the application in
BD cosmology gives us new insight to the expansion of the universe and the
 effect of BD scalar field on the subject.

\section{The model with two interacting fluids }

We start with the BD action for FRW universes filled with two perfect
fluids with energy densities $\rho_1$ and $\rho_2$. The
Friedmann equation is given by
\begin{equation}\label{1}
    3H^2+3\frac{k}{a^2}+3H\frac{\dot{\varphi}}{\varphi}-\frac{\omega}{2}\frac{\dot{\varphi}^2}{\varphi^2}=\frac{\rho_1+\rho_2}{\varphi},
\end{equation}
where $k=-1,0,1$. We assume that the two perfect fluids interact
through the interaction term $Q$ according to
\begin{eqnarray}
\label{2} &&\dot{\rho}_1+3H(\rho_1+p_1)=Q ,\\
\label{3} &&\dot{\rho}_2+3H(\rho_2+p_2)=-Q.
\end{eqnarray}
Although the nature of the $Q$ term is not clear at all,
for $Q>0$ there exists a transfer of energy from fluid with $\rho_2$
to the fluid with $\rho_1$. Also for $Q=0$ there are two
non-interacting fluids, each separately satisfies the standard
conservation equation. In the following we study the dynamics of the
universe for closed, open and flat universes separately.

\subsection{Closed and open power-law interacting cosmologies}

Let us now consider FRW cosmological models with $k=-1,1$ filled with
 interacting matter sources which satisfy the barotropic equation of
 state, i.e.
 \begin{equation}\label{4}
    p_1=\gamma_1 \rho_1,\ \ \ p_2=\gamma_2\rho_2,
 \end{equation}
 where $\gamma_1$  and $\gamma_2$ are equation of state (EoS) parameters. We
 assume that the scale factor and the BD scalar field behave as $a(t)\propto t^\alpha$ and $\varphi(t)\propto a^n\propto t^{n\alpha}$,
  where $\alpha$ and $n$ are constants.
This implies that $H\propto \alpha t^{-1}$. By taking into account the
curvature term $3k/a^2$ of equation (\ref{1}), we conclude that
$\alpha=1$ in order to obtain energy density scales over scalar
field on the RHS of the equation in the same manner as the curvature
term in the LHS. Since $a\propto t$ has no
acceleration, the universe will either expand or collapse with
constant velocity.

Now, from equation (\ref{1}) and the resultant equation from the
addition of equations (\ref{2})  and (\ref{3}) we obtain
\begin{eqnarray}
  \label{5} \rho_{k1}(t)=\frac{(1+n+3\gamma_2)(3(1+k+n)-n^2\omega/2)}{3(\gamma_2-\gamma_1)}t^{n-2}, \\
  \label{6}
  \rho_{k2}(t)=-\frac{(1+n+3\gamma_1)(3(1+k+n)-n^2\omega/2)}{3(\gamma_2-\gamma_1)}t^{n-2}.
\end{eqnarray}
The interacting term is also found to be,
\begin{equation}\label{7}
    Q(t)=-\frac{(1+n+3\gamma_1)(1+n+3\gamma_2)(3(1+k+n)-n^2\omega/2)}{3(\gamma_2-\gamma_1)}t^{n-3},
\end{equation}
which may be rewritten as
\begin{equation}\label{8}
    Q(t)=-(1+n+3\gamma_1)H\rho_{k1}=(1+n+3\gamma_2)H\rho_{k2}.
\end{equation}
This implies that it is proportional to the
expansion rate of the universe and to one of the individual
densities, so $Q\sim t^{n-3}$.

For the condition $n^2\omega < 6(1+n+k)$, and by defining
$\tau^{n-2}=(3(1+n+k)-n^2\omega/2)t^{n-2}$, the densities  (\ref{5}) and  (\ref{5}) become,
\begin{eqnarray}
\label{9} \rho_{k1}(\tau)=\frac{(1+n+3\gamma_2)}{3(\gamma_2-\gamma_1)}\tau^{n-2}, \\
  \label{10}
  \rho_{k2}(\tau)=-\frac{(1+n+3\gamma_1)}{3(\gamma_2-\gamma_1)}\tau^{n-2}.
\end{eqnarray}
Alternatively for $n^2\omega > 6(1+n+k)$ and taking
$\tau^{n-2}=-(3(1+n+k)-n^2\omega/2)t^{n-2}$, we have
\begin{eqnarray}
\label{11} \rho_{k1}(\tau)=-\frac{(1+n+3\gamma_2)}{3(\gamma_2-\gamma_1)}\tau^{n-2}, \\
  \label{12}
  \rho_{k2}(\tau)=+\frac{(1+n+3\gamma_1)}{3(\gamma_2-\gamma_1)}\tau^{n-2}.
\end{eqnarray}
As can be seen, in figure (1a), if we assume that one of the fluids is dust, for different values of $n$, for a particular $\gamma_1$ where
$\rho_{k1}=\rho_{k2}$, non of the fluid dominate. Before the crossing
point $\rho_{k2}$ dominates and after it $\rho_{k1}$ dominates. As
$n$ decreases, the domination of $\rho_{k1}$ over $\rho_{k2}$
occurs at larger value of negative $\gamma_1$ while $\gamma_2$ represent dust.
Also for those values of $n$ that $\rho_{k2}<0$, the Weak Energy Condition (WEC) ($\rho_{k1}\geq0 ,\rho_{k2}\geq0$) is not
satisfied and there exist no physical interpretation. In figure (1b) where
$\gamma_2=1/3$, we have the same argument and the only difference
 is that the curves are shifted smoothly to the right.

\begin{figure*}[t]
\centerline{
\includegraphics[width=8cm,height=8cm,angle=0]{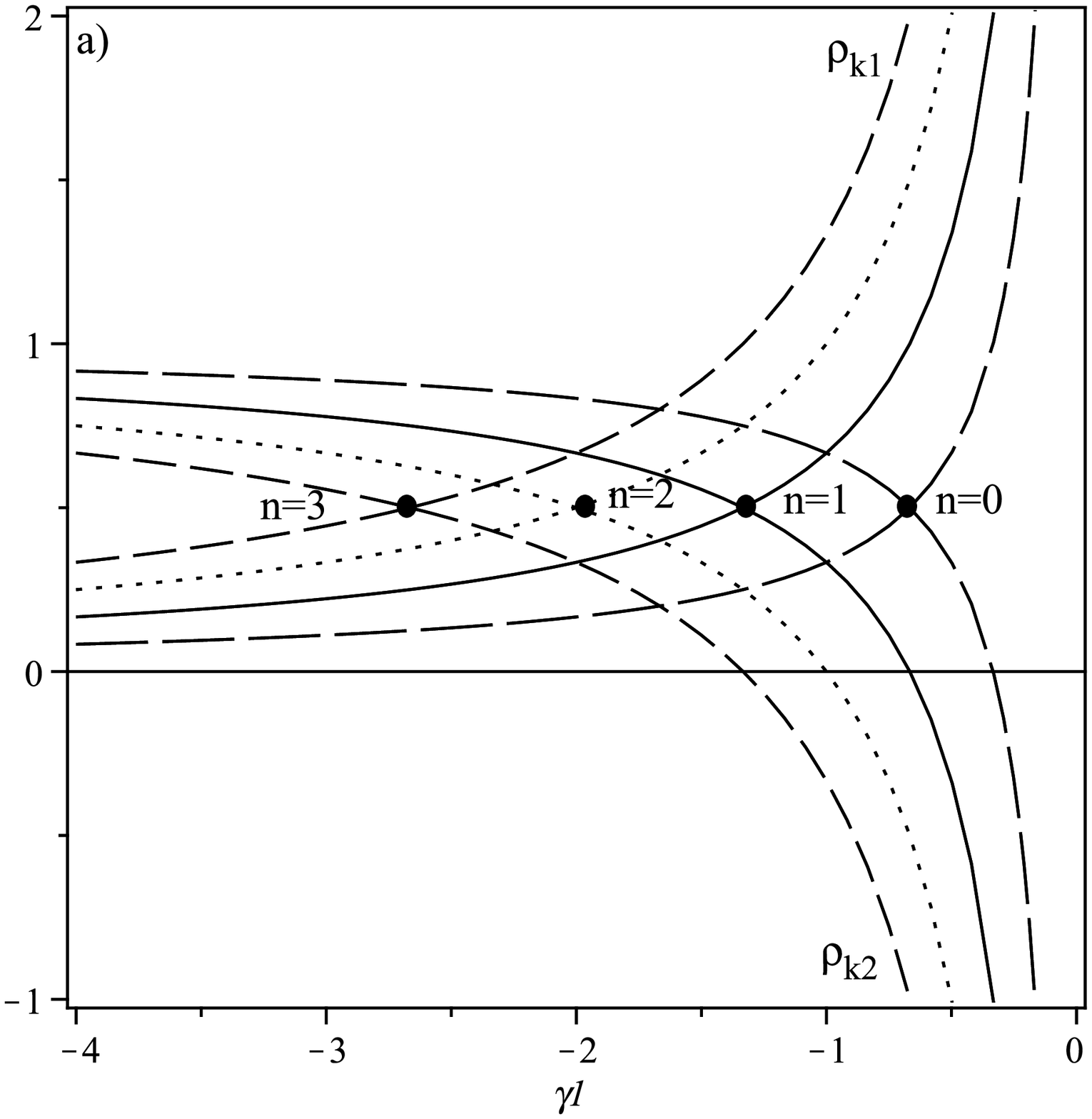}
\includegraphics[width=8cm,height=8cm,angle=0]{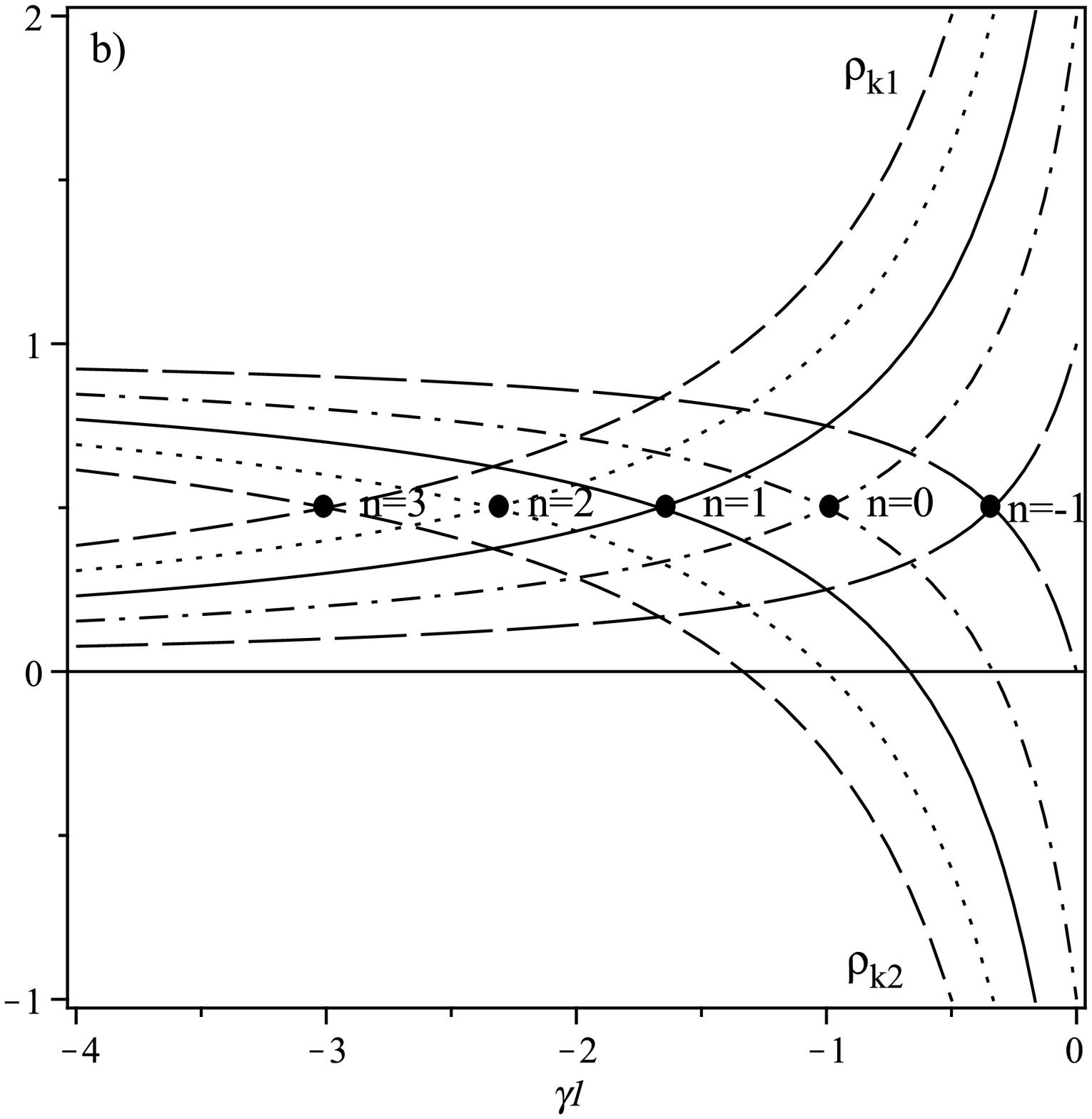}
}
\caption{ Plots of $\rho_{k1}\tau^{2-n}$ and
$\rho_{k2}\tau^{2-n}$ as functions of $\gamma_1$ with the condition $n^2\omega<6(n+2)$ for a) $\gamma_2=0$, b)$\gamma_2=1/3$
.}
\end{figure*}

Also, for the condition $n^2\omega>6(n+2)$ as shown in figures (2a) and (2b), the WEC is not satisfied and therefore
these is no physical interpretation for this case.

\begin{figure*}[t]
\centerline{
\includegraphics[width=8cm,height=8cm,angle=0]{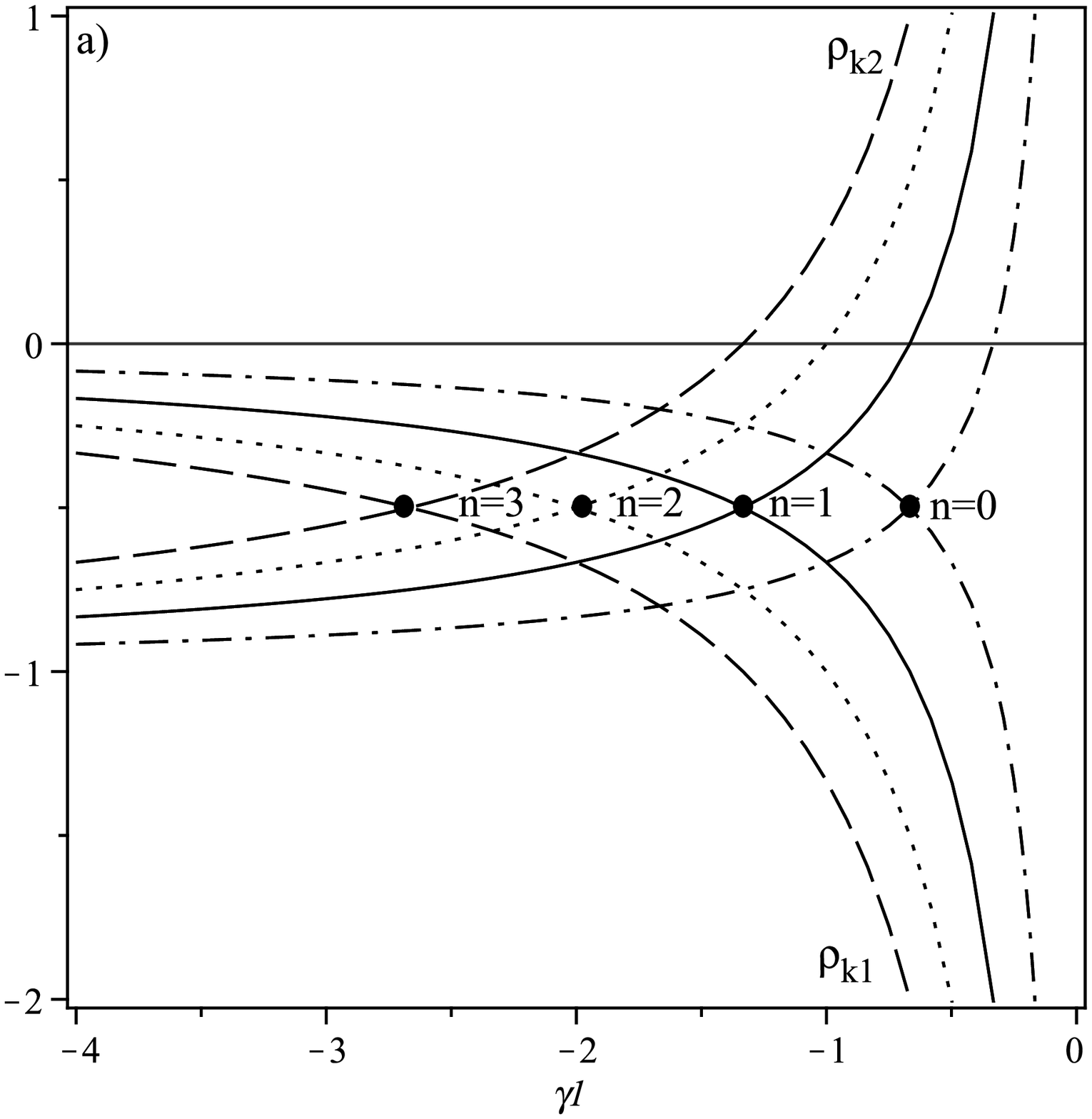}
\includegraphics[width=8cm,height=8cm,angle=0]{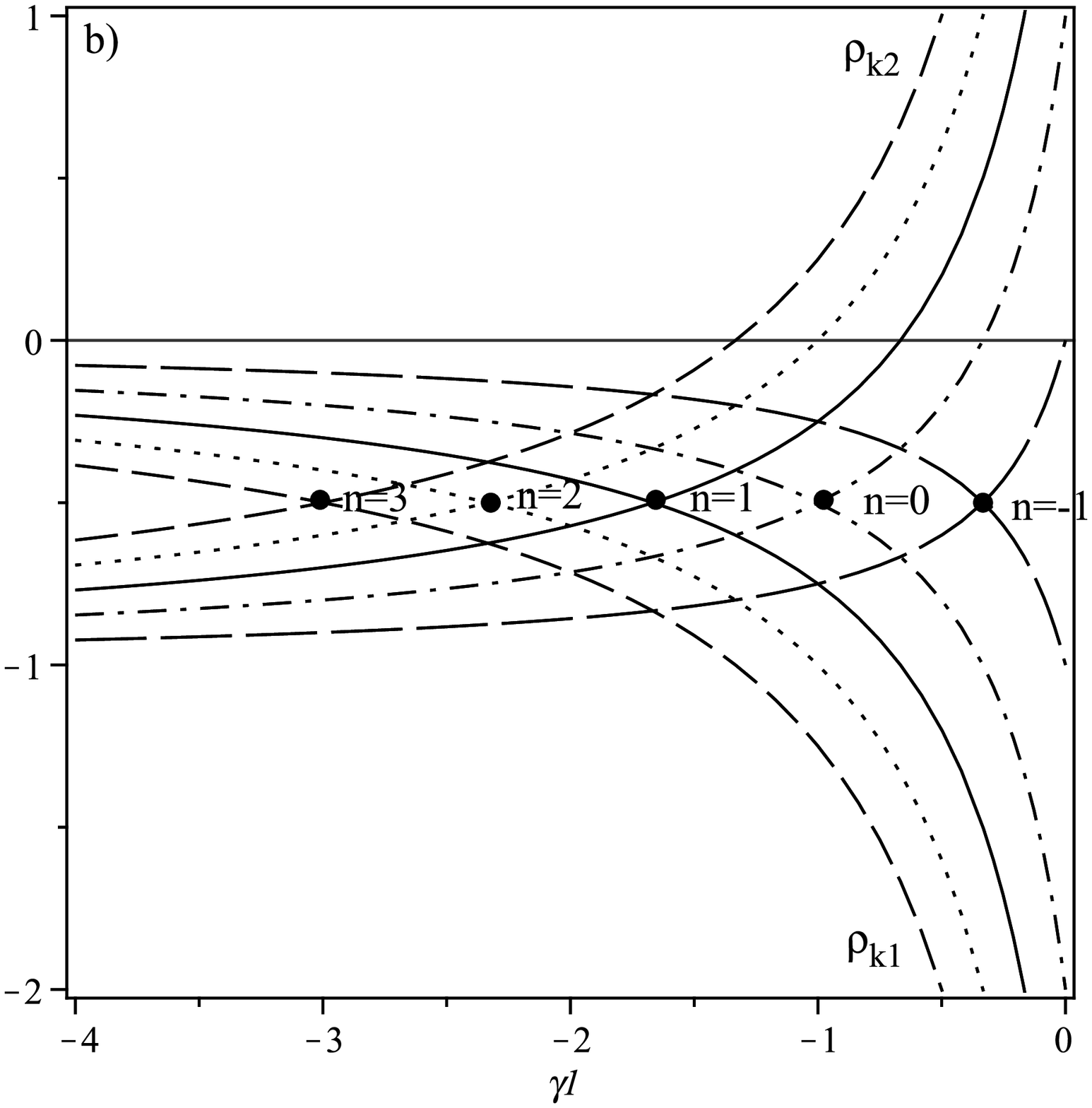}
}
\caption{ Plots of $\rho_{k1}\tau^{2-n}$ and
$\rho_{k2}\tau^{2-n}$ as functions of $\gamma_1$  with the condition $n^2\omega>6(n+2)$ for a) $\gamma_2=0$, b)$\gamma_2=1/3$
.}
\end{figure*}

From WEC and supposing that
$\gamma_2>\gamma_1$, and for the condition $n^2\omega<6(1+k+n)$, we have
\begin{eqnarray}
 \label{13} \gamma_2>-(n+1)/3,\ \ \gamma_1<-(n+1)/3.
\end{eqnarray}
Similarly for the condition $n^2\omega>6(1+k+n)$, we have
\begin{eqnarray}
 \label{14} \gamma_2<-(n+1)/3,\ \ \gamma_1>-(n+1)/3.
\end{eqnarray}
From these expressions we conclude that for $n>0$, always one of
the interacting fluids must be either a dark or a phantom fluid. For
example the constraints (\ref{13}) on the EoS parameters imply
that $Q > 0$, so the energy is transferred from a fluid with ($-1\geq
\gamma_2\geq -(n+1)/3$) or a phantom ($\gamma_1< -1$) fluid to the
dark component whose EoS parameter is $\gamma_2\geq
-(n+1)/3$. Also the constraint (\ref{14}) on the EoS parameters
imply that $Q < 0$, so the energy is transferred from a matter
component whose EoS parameter is $\gamma_2>-(n+1)/3$ to a dark
($-1\geq \gamma_1\geq -(n+1)/3$) or a phantom ($\gamma_1< -1$)
fluid.

The constant ratio of energies defined by $r_k=\rho_{k2}/\rho_{k1}=-\frac{1+n+3\gamma_1}{1+n+3\gamma_2}$
is another important factor that we consider in here. As can be seen, it is a $k$ independent function of the model EoS parameters $\gamma_1$, $\gamma_2$ and $n$. For cosmological scenarios which satisfy the requirement (\ref{13}) , we have $r_k=\rho_{k2}/\rho_{k1}>1$. Thus, $\rho_{k2}$ dominates over
 $\rho_{k1}$. On the other hand, for those which satisfy (\ref{14}), we have $r_k=\rho_{k2}/\rho_{k1}<1$ and $\rho_{k1}$ dominates over $\rho_{k2}$.

\subsection{Flat power-law interacting cosmologies}
From equations (\ref{1}), (\ref{2}), and (\ref{3}) for flat FRW
model we conclude that the general solution for energy distributions
are given by,
\begin{eqnarray}
  \label{15} \rho_{10}=\frac{\alpha(3(1+n)-n^2\omega/2)((n+3+3\gamma_2)\alpha-2)}{3(\gamma_2-\gamma_1)}t^{n\alpha-2}, \\
  \label{16}
  \rho_{20}=-\frac{\alpha(3(1+n)-n^2\omega/2)((n+3+3\gamma_1)\alpha-2)}{3(\gamma_2-\gamma_1)}t^{n\alpha-2}.
\end{eqnarray}
The interacting $Q$-term takes the form
\begin{equation}\label{17}
    Q(t)=\frac{\alpha(3(1+n)-n^2\omega/2)((n+3+3\gamma_1)\alpha-2)((n+3+3\gamma_2)\alpha-2)}{3(\gamma_2-\gamma_1)}t^{n\alpha-3},
\end{equation}
or in terms of the expansion rate of the universe and to one of the
individual densities,
\begin{equation}\label{18}
Q=\frac{((n+3+3\gamma_1)\alpha-2)}{\alpha}H\rho_{10}=-\frac{((n+3+3\gamma_2)\alpha-2)}{\alpha}H\rho_{20}.
\end{equation}
From equation (\ref{17}) we conclude that $Q\sim t^{n\alpha-3}$. For the
condition $n^2\omega < 6(1+n)$, and by definition
$\tau^{n\alpha-2}=(3(1+n)-n^2\omega/2)t^{n\alpha-2}$ one gets
\begin{eqnarray}
\label{19} \rho_{10}(\tau)=\frac{\alpha[(3+n+3\gamma_2)\alpha-2]}{3(\gamma_2-\gamma_1)}\tau^{n\alpha-2}, \\
  \label{20}
  \rho_{20}(\tau)=-\frac{\alpha[(3+n+3\gamma_1)\alpha-2]}{3(\gamma_2-\gamma_1)}\tau^{n\alpha-2}.
\end{eqnarray}
Alternatively, for the condition $n^2\omega > 6(1+n)$, and
$\tau^{n\alpha-2}=-(3(1+n)-n^2\omega/2)t^{n\alpha-2}$ we obtain
\begin{eqnarray}
\label{21} \rho_{10}(\tau)=-\frac{\alpha[(3+n+3\gamma_2)\alpha-2]}{3(\gamma_2-\gamma_1)}\tau^{n\alpha-2}, \\
  \label{22}
  \rho_{20}(\tau)=\frac{\alpha[(3+n+3\gamma_1)\alpha-2]}{3(\gamma_2-\gamma_1)}\tau^{n\alpha-2}.
\end{eqnarray}
From WEC, for $\gamma_2>\gamma_1$ and the condition $n^2\omega<6(1+n)$, we have
\begin{eqnarray}
   \label{23}
   \frac{2}{n+3+3\gamma_2}<\alpha<\frac{2}{n+3+3\gamma_1},\hspace{1.5cm} (\gamma_2>\gamma_1>-\frac{(n+3)}{3})\\
   \label{24} -\infty<\alpha<\frac{2}{n+3+3\gamma_1}, \
   \frac{2}{n+3+3\gamma_2}<\alpha<\infty, \hspace{1mm} (\gamma_2>-\frac{(n+3)}{3}>\gamma_1)\\
\label{25}
   \frac{2}{n+3+3\gamma_2}<\alpha<\frac{2}{n+3+3\gamma_1}. \hspace{1.5cm} (\gamma_1<\gamma_2<-\frac{(n+3)}{3})
 \end{eqnarray}
Also for $\gamma_2>\gamma_1$ and the condition $n^2\omega>6(1+n)$, we have
 \begin{eqnarray}
   \label{26}
   \frac{2}{n+3+3\gamma_1}<\alpha<\frac{2}{n+3+3\gamma_2},\hspace{1.5cm}(\gamma_2>\gamma_1>-\frac{(n+3)}{3})\\
   \label{27} -\infty,<\alpha<\frac{2}{n+3+3\gamma_2}, \
   \frac{2}{n+3+3\gamma_1}<\alpha<\infty,\hspace{1mm}(\gamma_2>-\frac{(n+3)}{3}>\gamma_1)\\
\label{28}
   \frac{2}{n+3+3\gamma_1}<\alpha<\frac{2}{n+3+3\gamma_2}.\hspace{1.5cm}(\gamma_1<\gamma_2<-\frac{(n+3)}{3})
 \end{eqnarray}
For $n>0$, equations (\ref{23}) and (\ref{26})  are valid for configurations
which include two interacting fluids obeying the dominant energy
condition (DEC), equations (\ref{24}) and (\ref{27})  are valid for
configurations where one interacting fluid obeys DEC and the other
is a phantom fluid, and equations (\ref{25}) and (\ref{28}) are
valid for the description of two interacting phantom fluids. Now let
us examine two interesting cases.

${\bf 1.}$ Dust-perfect fluid
interaction ($\gamma_1 = 0$, $\gamma_2 \neq 0$). For the condition
$n^2\omega<6(1+n)$, and from equations. (\ref{19}) and (\ref{20}), we have
\begin{eqnarray}
  \label{29} \rho_{10}&=&\frac{\alpha((n+3+3\gamma_2)\alpha-2)}{3\gamma_2}\tau^{n\alpha-2}, \\
   \label{30} \rho_{20}&=&\frac{-\alpha((n+3)\alpha-2)}{3\gamma_2}\tau^{n\alpha-2}.
\end{eqnarray}
Also, for the condition $n^2\omega>6(1+n)$, and from equations (\ref{21}) and (\ref{22}), we have
\begin{eqnarray}
  \label{31} \rho_{10}&=&-\frac{\alpha((n+3+3\gamma_2)\alpha-2)}{3\gamma_2}\tau^{n\alpha-2}, \\
   \label{32} \rho_{20}&=&\frac{\alpha((n+3)\alpha-2)}{3\gamma_2}\tau^{n\alpha-2}.
\end{eqnarray}
For the requirement of simultaneous fulfillment of the
conditions $\rho_{10}\geq0$ , $\rho_{20}\geq0$ and from equations
(\ref{23})-(\ref{25}) the following constraints must be satisfied for the case $n^2\omega<6(n+2)$
\begin{eqnarray}
   \label{33}
   \frac{2}{n+3+3\gamma_2}<\alpha<\frac{2}{n+3},\hspace{1.5cm}(\gamma_2>0>-\frac{(n+3)}{3})\\
   \label{34} -\infty<\alpha<\frac{2}{n+3},\
   \frac{2}{n+3+3\gamma_2}<\alpha<\infty, \hspace{1mm}(\gamma_2>-\frac{(n+3)}{3}>0)\\
\label{35}
   \frac{2}{n+3+3\gamma_2}<\alpha<\frac{2}{n+3}. \hspace{1.5cm}(0<\gamma_2<-\frac{(n+3)}{3})
 \end{eqnarray}
Also for $n^2\omega>6(1+n)$ and from equations
(\ref{26})-(\ref{28}) we obtain,
 \begin{eqnarray}
   \label{36}
   \frac{2}{n+3}<\alpha<\frac{2}{n+3+3\gamma_2},\hspace{1.5cm}(\gamma_2>0>-\frac{(n+3)}{3})\\
   \label{37} -\infty<\alpha<\frac{2}{n+3+3\gamma_2}, \
   \frac{2}{n+3}<\alpha<\infty , \hspace{1mm}(\gamma_2>-\frac{(n+3)}{3}>0\\
\label{38}
   \frac{2}{n+3}<\alpha<\frac{2}{n+3+3\gamma_2}.\hspace{1.5cm}(0<\gamma_2<-\frac{(n+3)}{3})
 \end{eqnarray}
As a specific example, we shall now consider in some detail the
dust-radiation interaction ($\gamma_1=0 , \gamma_2=1/3$). In this
case we have for $n^2\omega<6(1+n)$
\begin{eqnarray}
  \nonumber  \rho_{10}&=&\alpha[(n+4)\alpha-2]\tau^{n\alpha-2},\ \ \ \ \ p_1=0, \\
  \label{39} \rho_{20}&=&-\alpha[(n+3)\alpha-2]\tau^{n\alpha-2},\ \ p_2=\frac{1}{3}\rho_{20}.
\end{eqnarray}
For the condition $n^2\omega>6(1+n)$, we have,
\begin{eqnarray}
  \nonumber \rho_{10}&=&-\alpha[(n+4)\alpha-2]\tau^{n\alpha-2},\  \ p_1=0, \\
  \label{40} \rho_{20}&=&\alpha[(n+3)\alpha-2]\tau^{n\alpha-2},\ \ \ \  p_2=\frac{1}{3}\rho_{20}.
\end{eqnarray}
In figure (3a), where $n^2\omega<6(1+n)$,
 for different values of $n$, we first have radiation dominated and then dust dominated. Since the universe in both radiation and dust dominated has to be in
 deceleration phase, we expect that dust domination continues only till $\alpha <1$. In addition, from figure (3a), $n$ under no condition can be greater than $-1.5$, since $\alpha$ becomes greater than one and the universe starts to accelerate which contradicts the universe decelerating behaviour in radiation dominated and matter dominated era. Further, it can be seen that as $n$ decreases, the crossing of the dust and radiation distributions takes place at greater values. As shown in figure (3b), there is no physical solution for the case $n^2\omega>6(1+n)$.

\begin{figure*}[t]
\centerline{
\includegraphics[width=8cm,height=8cm,angle=0]{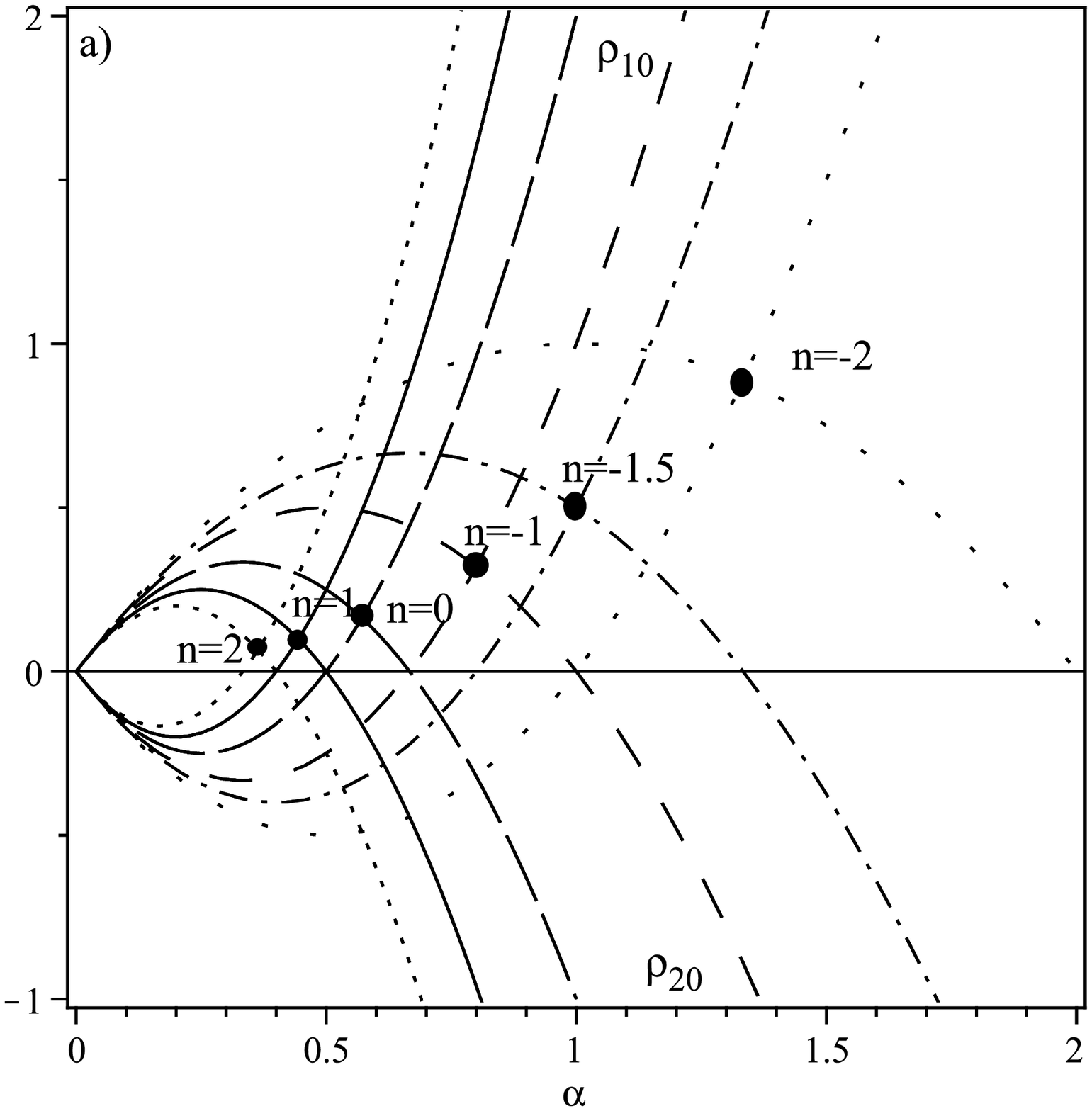}
\includegraphics[width=8cm,height=8cm,angle=0]{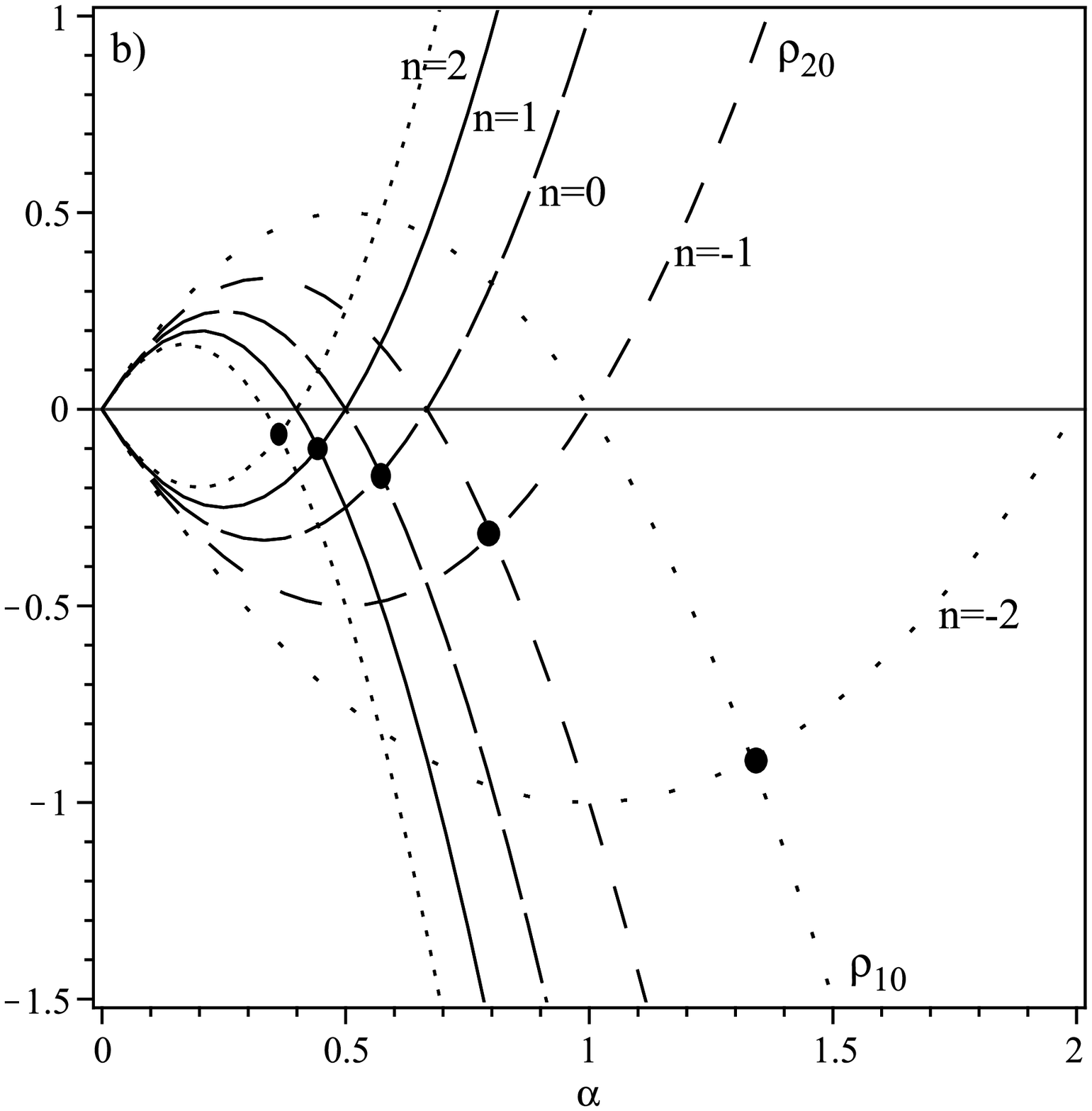}
}
\caption{ The behaviors of interacting dust distribution, $\rho_{10}$,
 and radiation distribution, $\rho_{20}$, are plotted as functions of $\alpha$
   for conditions a) $n^2\omega<6(1+n)$  and b) $n^2\omega>6(1+n)$. }
\end{figure*}

${\bf 2.}$ Phantom-perfect fluid interaction (say $\gamma_1=-4/3$,
$\gamma_2 \neq 0$). For the condition $n^2\omega<6(1+n)$, we have
\begin{eqnarray}
  \nonumber \rho_{10}&=&\frac{\alpha((n+3+3\gamma_2)\alpha-2)}{3\gamma_2+4}\tau^{n\alpha-2}, \\
   \label{41}
   \rho_{20}&=&\frac{-\alpha((n-1)\alpha-2)}{3\gamma_2+4}\tau^{n\alpha-2}.
\end{eqnarray}
For the condition $n^2\omega>6(1+n)$,
\begin{eqnarray}
  \nonumber \rho_{10}&=&-\frac{\alpha((n+3+3\gamma_2)\alpha-2)}{3\gamma_2+4}\tau^{n\alpha-2}, \\
   \label{42}
   \rho_{20}&=&\frac{\alpha((n-1)\alpha-2)}{3\gamma_2+4}\tau^{n\alpha-2}.
\end{eqnarray}

As a specific example we shall now consider in some detail the
phantom-dust interaction ($\gamma_1=-4/3 , \gamma_2=0$). In this
case we have for $n^2\omega<6(1+n)$,
\begin{eqnarray}
  \nonumber \rho_{10}&=&\frac{\alpha((n+3)\alpha-2)}{4}\tau^{n\alpha-2}, \\
   \label{41}
   \rho_{20}&=&\frac{-\alpha((n-1)\alpha-2)}{4}\tau^{n\alpha-2}.
\end{eqnarray}
For $n^2\omega>6(1+n)$,
\begin{eqnarray}
  \nonumber \rho_{10}&=&-\frac{\alpha((n+3)\alpha-2)}{4}\tau^{n\alpha-2}, \\
   \label{42}
   \rho_{20}&=&\frac{\alpha((n-1)\alpha-2)}{4}\tau^{n\alpha-2}.
\end{eqnarray}

\begin{figure*}[t]
\centerline{
\includegraphics[width=8cm,height=8cm,angle=0]{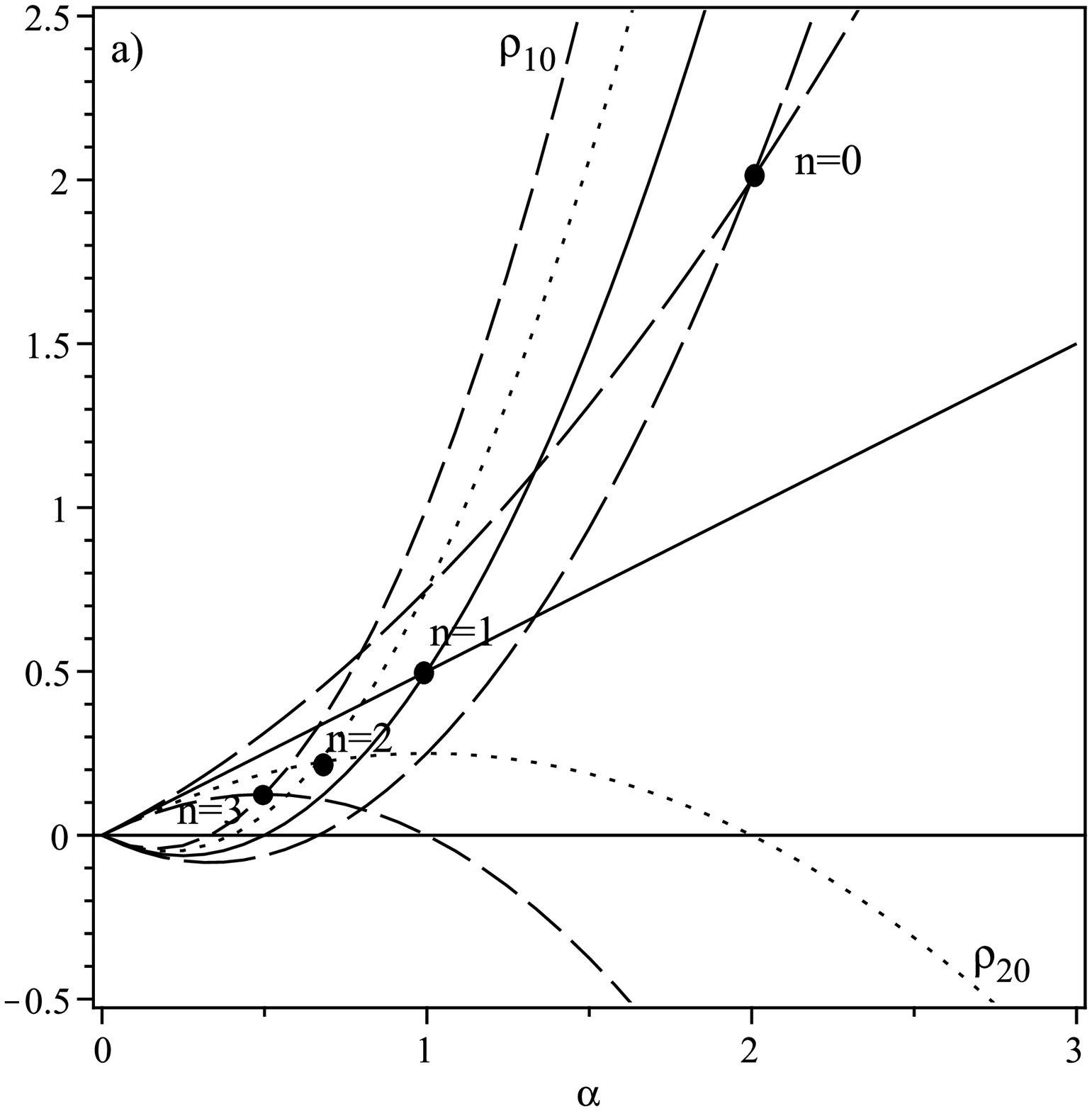}
\includegraphics[width=8cm,height=8cm,angle=0]{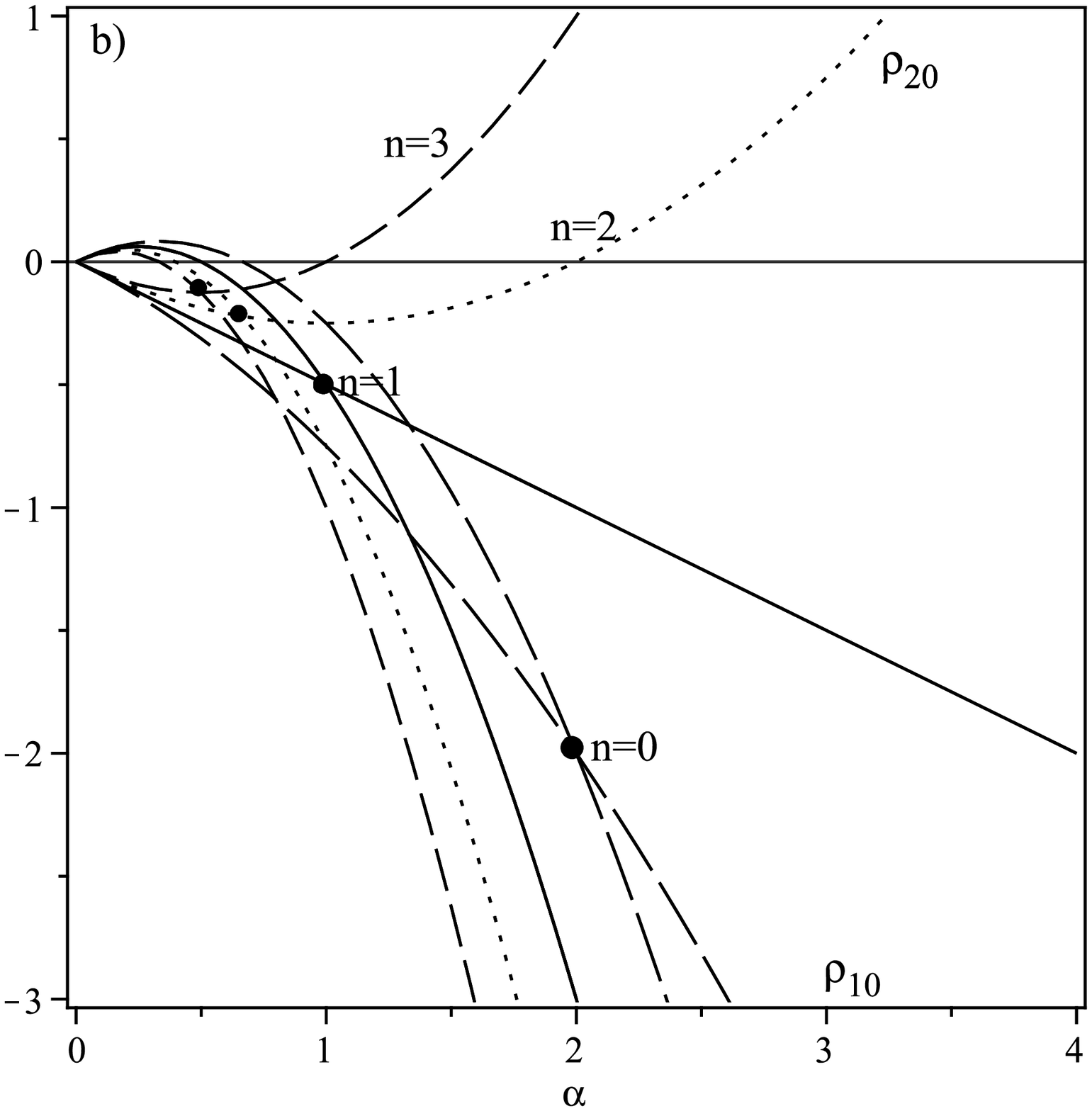}
}
\caption{The behaviors of interacting phantom distribution, $\rho_{10}$,
 and matter distribution, $\rho_{20}$, for conditions a) $n^2\omega<6(1+n)$ and b) $n^2\omega>6(1+n)$.}
\end{figure*}
From figure (4a), one can see that matter domination occurs before phantom domination. Further, it can be seen that as $n$ decreases, the crossing of the phantom and dust distributions takes place at greater values of $\alpha$. Moreover, for $\alpha=1$ when the universe starts accelerating and phantom dominates over matter, the value of $n$ is one. Figure (4b) shows no physical interpretation for the case $n^2\omega>6(1+n)$.

\section{The effective fluid interaction}
In this section we study the conditions under which the two
interacting sources are equivalent to an effective fluid
described as the interaction of two-fluid mixture. This can
be made by introducing an effective pressure $p_{eff}$, i.e.
\begin{equation}\label{37}
    p_1+p_2=\gamma_1\rho_1+\gamma_2\rho_2=p_{eff},
\end{equation}
which has an equation of state given by
\begin{equation}\label{38}
    p_{eff}=\gamma\rho_{eff}=\gamma(\rho_1+\rho_2),
\end{equation}
where $\gamma$ is the effective EoS parameter. Note that
the equation of state of associated effective fluid is not produced
by physical particles and their interaction[17]. Making some
algebraic manipulations with equations (\ref{37}) and (\ref{38}) we find
that the the effective EoS parameter, $\gamma$, is related to the
parameter $\alpha$ by,
\begin{equation}\label{39}
    \gamma=-1-\frac{n^2\omega+n^2-n}{3(n+1)-n^2\omega/2}+\frac{2+n}{\alpha[3(n+1)-n^2\omega/2]}.
\end{equation}
We also find the behavior of the BD parameter $\omega$ in terms of $n$ ($n\neq 0$)in:

{\it Inflation era} ($\alpha>> 1$, $\gamma=-1$): \ \ \ \ \ \ \ \ \ \ \ \ \ \ \ \ \ \ \ \  $ \omega=-1+\frac{1}{n}$ ${\bf .}$

{\it  Radiation dominated era} ($\alpha =1/2$, $\gamma=1/3$):  \ \ \ $\omega=-3-\frac{3}{n}$

 {\it Matter dominated era} ($\alpha =2/3$, $\gamma=0$): \ \ \  \ \ \ \ \ \ \  $\omega=-2-\frac{1}{n}$

{\it Dark energy dominated era} ($\alpha \backsimeq 1.1$, $\gamma=-1$):  \ $\omega=-1+\frac{2}{n}+\frac{2}{n^2}$

 It shows that in the inflation
and cosmological constant era the behaviour of $\omega$ in terms of
$n$ is different from in radiation and matter dominated era.
In case of $n =0$ ( or $\phi=\phi_0$)  eq.(\ref{39}) becomes,
\begin{equation}\label{40}
    \gamma=\frac{2-3\alpha}{3\alpha},
\end{equation}
 in compatible with the standard cosmology. That is, in radiation era where $\alpha=1/2$, the EoS parameter becomes $1/3$, in matter dominated era where
 $\alpha=2/3$, the EoS parameter becomes zero, in inflation era where $\alpha \rightarrow \infty$, the EoS parameter becomes $-1$ and in the current dark energy era where $\alpha$ is just greater than one, the EoS parameter becomes about $-1/3$ which is compatible with the observational data.

\section{Discussion}

In this paper we have provided a detailed description
for power law scaling cosmological models in the case of a
FRW universe in BD theory dominated by two interacting perfect fluid
components during the expansion. We have shown that
in this mathematical description it is possible for each
fluid component to require that the WEC may be simultaneously fulfilled in order to have
reasonable physical values of EoS parameters in the formulation.
So from the required conditions we may
gain some insights for understanding essential features
of two fluid interactions in power law FRW BD cosmologies.
In closed  or open universes, if one of the fluids is dust or radiation, the second fluid
has to be dark energy or phantom. As the power of BD scalar field, $n$, a function of time, decreases, the second fluid  changes its behaviour from dark energy to phantom. If one of the fluids is radiation, the change of behaviour of the second fluid, as $n$ decreases, is slower in comparison to the dust fluid.

 In flat universes, if the two fluids are dust and radiation, their behaviuor as a function of scale factor power of time, $\alpha$, changes with $n$. It has been shown that as $n$ decreases, the domination of dust over radiation occurs for smaller values of $\alpha$. It also gives us a constraint on $n$  not to be greater than values that makes the universe accelerate. Similarly, if the two fluids are phantom and dust, again their behaviour as a function of  $\alpha$ changes with $n$. Firstly, for a particular $n$, the domination of dust occurs before the domination of phantom. Secondly, as $n$ decreases the domination of phantom occurs at greater values of $\alpha$. Thirdly, $n$ is constrained such that in dust dominated era $\alpha$ has to be less than one for a decelerated universe.

 Finally, we investigate the effective fluid interaction in the model. We find an expression for the effective EoS parameter, $\gamma$, as a function of $\alpha$, $n$ and $\omega$. The behaviour of the BD parameter, $\omega$, as a function of $n$ in different epoch of the universe expansion is given. It has been found that $\omega$ is inversely (negatively) proportional to $n$ in radiation and matter dominated era, whereas inversely (positively) proportional to $n$ in inflation and dark energy dominated era. For the constant scalar field the effective EoS parameter turns out to be the one expected in standard cosmology.
 
 {\bf Acknowledgement}

The authors would like to thank the anonymous reviewer for the careful review and helpful comments. We would also like to thank University of Guilan, Research Council for their support.

\end{document}